# Electrochemical conversions in a microfluidic chip for xenobiotic metabolism and proteomics

*Albert van den Berg - University of Twente*

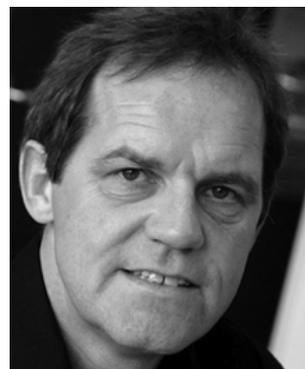

## Biography

*Albert van den Berg received his MSc in applied physics in 1983, and his PhD in 1988 both at the University of Twente, the Netherlands. From 1988-1993 he worked in Neuchatel, Switzerland, at the CSEM and the University (IMT) on miniaturized chemical sensors. In 1998 he was appointed as part-time professor "Biochemical Analysis Systems", and later in 2000 as full professor on Miniaturized Systems for (Bio)Chemical Analysis in the faculty of Electrical Engineering and part of the MESA+ Institute for Nanotechnology. In 1994 he initiated together with Prof. Bergveld the international MicroTAS conference series. He published over 400 peer reviewed publications (H=53) a.o. in Science, Nature, PNAS, NanoLetters etc. He received several honors and awards such as Simon Stevin (2002), two ERC Advanced (2008, 2015) and three ERC Proof of Concept (2011, 2013, 2016) grants, Simon Stevin award (engineering sciences), Spinoza prize (2009), Distinguished University Professor (Twente, 2010), Distinguished Professor (South China Normal University SNCU, 2012) and board member of the Royal Dutch Academy of Sciences (KNAW) (2011-2016). In 2014 he was appointed scientific director of the MIRA institute for Biomedical Engineering. In 2017 he became co-PI of the Max Planck – University of Twente Center for Complex Fluid Dynamics.*

## Introduction

The ability to generate xenobiotic metabolites in an entirely instrumental fashion is of great use for pharmaceutical research and toxicological testing. At the electrode surface in a microfluidic electrochemical cell, conversion efficiencies are increased, time between electrochemical generation of metabolites and their detection is reduced and microfabrication techniques enable the integration of multiple reactors (e.g., an electrochemical cell and a micromixer) in a single device. These same devices can also be used to electrochemically cleave proteins, thereby generating fragments that allows one to identify the protein by means of mass spectrometric detection and database searching.

To this end, we developed a microfluidic electrochemical cell with integrated boron doped diamond electrode and a volume of ~160 nL (Figure 1A). Compared to, e.g., platinum, this superior electrode material has a reduced proneness to fouling by organic substances such as proteins. Coupling of microfluidic electrochemical cells in an on-line electrochemistry-mass spectrometry (EC-MS) arrangement using an electrospray ionization (ESI) interface (Figure 1B) allows one to rapidly detect electrogenerated products, including short-lived and potentially toxic drug metabolites that are too unstable to be detected by conventional methods.[1] This enables one to develop a powerful instrumental alternative to established in vitro drug metabolism tests that aim to mimic metabolic certain activity of enzymes from the cytochrome P450 family.[2]

## Reactive metabolites

The use of these devices for mimicking xenobiotic metabolism was demonstrated by electrochemical oxidation of 1-hydroxypyrene (1-OHP). Electrogenerated metabolites were allowed to react immediately (delay between 0.1-1.5 s) with haemoglobin (Hb) in a micromixer that was located just 1.5 mm after the working electrode (WE) in of the electrochemical cell. Reactivity of these metabolites was evaluated by mass spectrometric detection of the reaction products coming out of the mixer, following separation by liquid chromatography.[3] Deconvoluted mass spectra in Figure 2 show that both the α and β subunits of Hb are modified by pyrene quinone, one of the known metabolites of 1-OHP.

## Protein cleavage

Mass spectrometric analysis of intact proteins is only possible with the most advanced instrument, so there is a need to cleave proteins at well-defined locations to produce small fragments that can be detected. As an alternative to chemical cleavage methods, electrochemical peptide bond oxidation followed by cleavage has shown to be possible specifically adjacent to tyrosine and tryptophan residues at positive potentials.[4,5] In addition, at negative potentials, disulfide bonds can be reduced and cleaved.[6] These possibilities were demonstrated using insulin as a model compound, which has four tyrosine residues and three disulfide bonds (see figure 3).

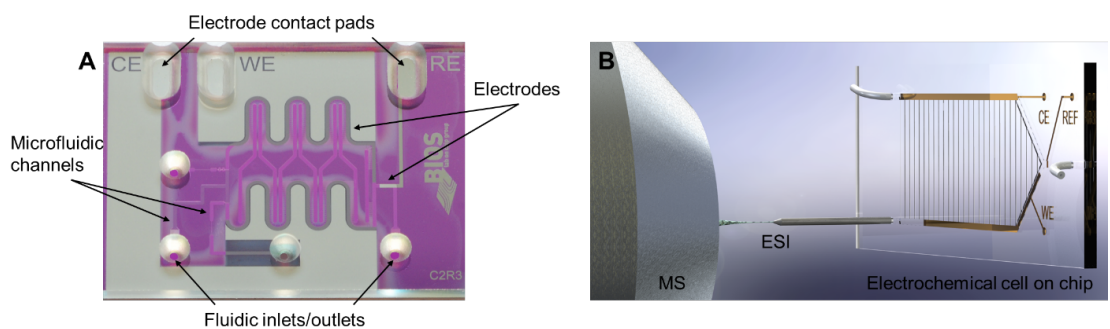

*Figure 1. A: Example of a microfluidic chip for electrochemical conversion (photo by Henk van Wolferen). B: Artist impression of an EC-MS arrangement with a microfluidic electrochemical cell. Adapted with permission from ref. 1. Copyright 2015 American Chemical Society.*

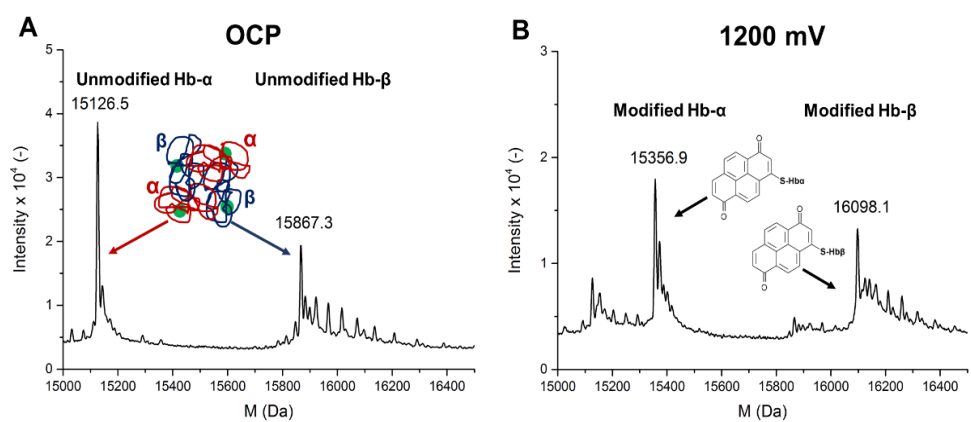

*Figure 2. Deconvoluted mass spectra of haemoglobin, showing both subunits (α: 15126.5 Da and β: 15867.3 Da) in the presence of 1-OHP. A: The WE potential was set to open circuit potential (OCP) in the control measurement. B: The WE potential was set to 1200 mV. Electrochemical oxidation of 1-OHP leads to modification of both subunits, which became evident from the appearance of new peaks at 15356.9 Da and 16098.1 Da. This represents mass shifts of 230.4 and 230.8 m/z, which can be related to the addition of one pyrene quinone. Reproduced from ref. 3 by permission of The Royal Society of Chemistry.*

## Conclusions

Electrochemistry on chip has the potential to play an important role in a wide range of analytical applications. We gave two examples, the first being xenobiotic metabolism, which can relate to drug screening or the evaluation of the toxicity of environmental pollutants. Reactive metabolites can be electrochemically generated and evaluated for their potential toxicity in a completely integrated microfluidic platform consisting of both an electrochemical and chemical microreactor. Furthermore, in the field of proteomics one can benefit from these devices to cleave proteins for their identification and characterization. This entirely instrumental approach is faster than enzymatic procedures and can be combined on-line with analytical instruments such as mass spectrometers. With improved control over reaction conditions, reduced sample handling and faster analysis the impact of microfluidics can be expected to become more extensive in these and other areas.

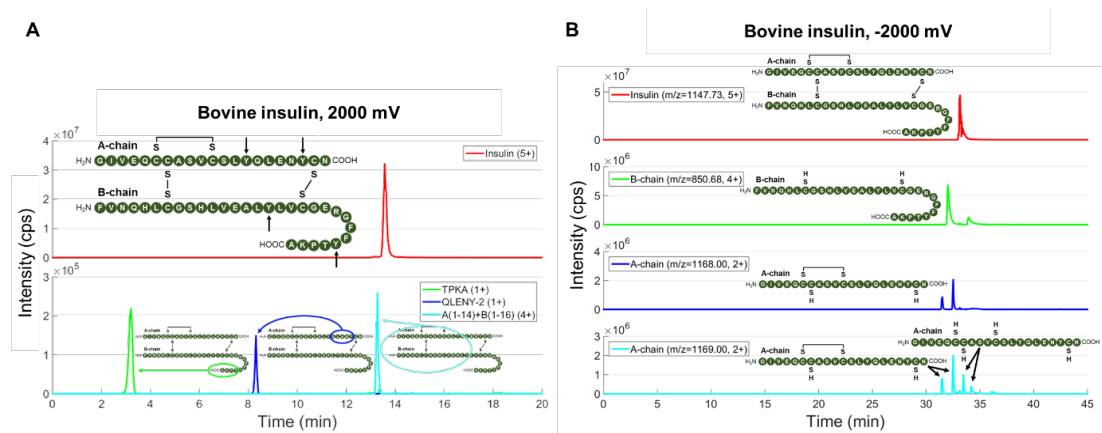

Figure 3. Electrochemical cleavage of insulin, which contains four tyrosine residues and three disulfide bonds. A: Insulin was electrochemically cleaved at 2000 mV. Extracted ion chromatograms are shown for intact insulin and three cleavage products, demonstrating that cleavage occurs at all four tyrosine residues (indicated with arrows in the inset). Adapted with permission from ref. 5. Copyright 2016 American Chemical Society. B: Disulfide bonds were reduced at -2000 mV. Extracted ion chromatograms are shown for intact insulin, the separated B-chain and the separated A-chain, showing that both the intermolecular and the intramolecular disulfide bonds were reduced.

## References

1. F.T.G. van den Brink, L. Büter, M. Odijk, W. Olthuis, U. Karst and A. van den Berg, *Anal. Chem.*, 2015, **87**, 1527-1535.
2. A. Baumann, U. Karst, *Expert Opin. Drug Metab. Toxicol.*, 2010, **6**, 715-731.
3. F.T.G. van den Brink, T. Wigger, L. Ma, M. Odijk, W. Olthuis, U. Karst and A. van den Berg, *Lab Chip*, 2016, **16**, 3990-4001.
4. H.P. Permentier, A.P. Bruins, *J. Am. Soc. Mass. Spectrom.*, 2004, **15**, 1707-1716.
5. F.T.G. van den Brink, T. Zhang, L. Ma, J. Bomer, M. Odijk, W. Olthuis, H.P. Permentier, R.P.H. Bischoff and A. van den Berg, *Anal. Chem.*, 2016, **88**, 9190-9198.
6. A. Kraj, H. Brouwer, N. Reinhoud and J.-P Chervet, *Anal. Bioanal. Chem.*, 2013, **405**, 9311-9320.